\documentclass[aps,showpacs,superscriptaddress,
preprintnumbers,onecolumn]{revtex4}%
\usepackage{amsfonts}
\usepackage{amsmath}
\usepackage{hyperref}
\usepackage{scalefnt}
\usepackage{amssymb}
\usepackage{graphicx,pstricks}
\usepackage{graphicx}%
\topmargin
20pt
\newtheorem{theorem}{Theorem}

\newtheorem{lemma}[theorem]{Lemma}
\newtheorem{proposition}[theorem]{Proposition}
\newenvironment{proof}[1][Proof]{\noindent\textbf{#1.} }{\ \rule{0.5em}{0.5em}}

\begin{document}
\title{A generalization of boson normal ordering}
\author{Toufik Mansour}
\email{toufik@math.haifa.ac.il}
\affiliation{Center for Combinatorics, LPMC, Nankai University, 300071 Tianjin, P.R. China}
\author{Matthias Schork}
\email{mschork@member.ams.org}
\affiliation{Alexanderstr. 76, 60489 Frankfurt, Germany}
\author{Simone Severini}
\email{ss54@york.ac.uk}
\affiliation{Department of Mathematics and Department of Computer Science, University of
York, Heslington, YO10 5DD York, U.K.}

\begin{abstract}
In this paper we define generalizations of boson normal ordering. These are
based on the number of contractions whose vertices are next to each other in
the linear representation of the boson operator function. Our main motivation
is to shed further light onto the combinatorics arising from algebraic and
Fock space properties of boson operators.

\end{abstract}
\pacs{02.10.Ox}
\maketitle

\section{Introduction}

Generalizations and refinements of normal ordering have been studied in the
literature (see \cite{Kat92, thesis, Kat02, p}). Here we introduce a
generalization of the normal ordering for boson operator functions. The
generalization is based on the number of contractions whose vertices are next
to each other in the linear representation of the function itself. In this
way, we introduce a parameter associated to these specific contractions. The
use of the parameter allows to refine the set of contractions and to give a
statistics on this set, obtaining the standard normal ordering as a special
case. Additionally, we outline a framework for an even more general approach
to normal orderings, where contractions between vertices of different
distances are allowed to have arbitrary weights. With this tools we are able
to extend some of the results of Katriel \cite{Kat02}, where a concise theory
of Bell polynomials was described in terms of the algebraic and Fock space
properties of the boson operators. Keeping in mind that general formulas for
normally ordered forms are commonly difficult to obtain (see \cite{bl}), our
analysis is potentially useful in computing the standard normal ordering,
since it provides a larger perspective, at least from the combinatorial point
of view.

The paper is organized as follows. In Section II we introduce a generalization
of the boson normal ordering. In Section III we discuss some physical aspects
of this generalization. In Section IV we give an explicit formula for the
generalized normally ordered form of the function $(a^{\dagger}a)^{n}$
(Theorem \ref{mth2}). In order to establish the result, we prove that there
exists a bijection between the set of contractions of $(a^{\dagger}a)^{n}$
having degree $k$ and the set of partitions of $n$ with $k$ rises (Proposition
\ref{pro3}). Considering coherent states (see, \emph{e.g.}, \cite{k}), we
prove relations between the generalized normal ordering and integer sequences
like the Bell numbers and Stirling numbers, thus generalizing observations of
\cite{Kat02}. Finally, in Section V we describe a rather general framework for
generalizations of normal ordering by introducing weights for contractions
satisfying some natural conditions. However, on this general level we neither
consider explicit expressions nor physical applications; this will be left for
future investigations.

\section{A generalization of the boson normal ordering}

Let $b$ and $b^{\dagger}$ be the boson annihilation and creation operators
satisfying the commutation relation
\begin{equation}
\label{comm}[b,b^{\dagger}]\equiv bb^{\dagger}-b^{\dagger}b=I.
\end{equation}
The \emph{normal ordering} is a functional representation of boson operator
functions in which all the creation operators stand to the left of the
annihilation operators. A function $F(b,b^{\dagger})$ can be seen as a word
(of possible infinite length) on the alphabet $\{b,b^{\dagger}\}$. We denote
by $\mathcal{N}[F(b,b^{\dagger})]$ the normal ordering of a function
$F(b,b^{\dagger})$. We can obtain $\mathcal{N}[F(b,b^{\dagger})]$ from
$F(b,b^{\dagger})$ by means of contractions and double dot operations. The
\emph{double dot operation} deletes all the letters $\varnothing$ and
$\varnothing^{\dagger}$ in the word and then arranges it such that all the
letters $b^{\dagger}$ precede the letters $b$. For example, $:b^{k}%
(b^{\dagger})^{l}:$ $=(b^{\dagger})^{l}b^{k}$. A \emph{contraction} consists
of substituting $b=\varnothing$ and $b^{\dagger}=\varnothing^{\dagger}$ in the
word whenever $b$ precedes $b^{\dagger}$. Among all possible contractions, we
also include the null contraction, that is the contraction leaving the word as
it is. Specifically,
\begin{equation}
\label{normal}\mathcal{N}[F(b,b^{\dagger})]=F(b,b^{\dagger})=\sum:\{\text{all
possible contractions}\}:.
\end{equation}
For example, the normal ordering of the word $bb^{\dagger}bbb^{\dagger}bb$ is
$(b^{\dagger})^{2}b^{5}+4b^{\dagger}b^{4}+2b^{3}$.

Several authors established different connections between Stirling and Bell
numbers and normally ordered forms (for example, see~\cite{Kat02, thesis} and
references therein). Consider the number operator $N=b^{\dagger}b$. The
normally ordered form of its $n$-th power can be written as
\begin{equation}
\left(  b^{\dagger}b\right)  ^{n}=\sum_{k=1}^{n}S(n,k)(b^{\dagger})^{k}b^{k},
\label{eqkat}%
\end{equation}
where the integers $S(n,k)$ are the so called Stirling numbers of second kind
(see \cite[Seq. A008277]{Int}). One may also define the so called Bell
polynomials $B(n,x)=\sum_{k=1}^{n}S(n,k)x^{k}$ and Bell numbers
$B(n)=B(n,1)=\sum_{k=1}^{n}S(n,k)$. The Stirling numbers satisfy the following
recurrence relation
\begin{equation}
\label{Stirl}S(n+1,k)=kS(n,k)+S(n,k-1)
\end{equation}
with the initial conditions $S(n,0)=\delta_{n,0}$ and $S(n,k)=0$ for $k>n$
\cite{Int}.

\begin{figure}[h]
\begin{center}
\begin{pspicture}(0,0)(10,.4)
\setlength{\unitlength}{3mm} \linethickness{0.3pt}
\multips(0,0)(3,0){4}{\pscircle(0,0){.2}\pscircle(.5,0){.2}\pscircle*(1,0){.2}\pscircle*(1.5,0){.2}}
\linethickness{0.8pt}\qbezier(10,.1)(11.6,1)(13.2,.1)
\qbezier(20,.1)(22.5,1)(25,.1)\qbezier(31.6,.1)(32.4,1)(33.2,.1)
\put(-.2,-1.2){$4$}\put(1.5,-1.2){$3$}\put(3.2,-1.2){$2$}\put(4.9,-1.2){$1$}
\put(9.7,-1.2){$4$}\put(11.4,-1.2){$3$}\put(13.1,-1.2){$2$}\put(14.8,-1.2){$1$}
\put(19.7,-1.2){$4$}\put(21.4,-1.2){$3$}\put(23.1,-1.2){$2$}\put(24.8,-1.2){$1$}
\put(29.7,-1.2){$4$}\put(31.4,-1.2){$3$}\put(33.1,-1.2){$2$}\put(34.8,-1.2){$1$}
\end{pspicture}
\par
\begin{pspicture}(0,0)(10,.7)
\setlength{\unitlength}{3mm} \linethickness{0.3pt}
\multips(0,0)(3,0){3}{\pscircle(0,0){.2}\pscircle(.5,0){.2}\pscircle*(1,0){.2}\pscircle*(1.5,0){.2}}
\linethickness{0.8pt}\qbezier(1.6,.1)(3.3,1.2)(5,.1)
\qbezier(10,.1)(12.5,1.2)(15,.1)\qbezier(11.6,.1)(12.4,.8)(13.2,.1)
\qbezier(20,.1)(21.65,1.2)(23.3,.1)\qbezier(21.6,.1)(23.25,1.2)(24.9,.1)
\put(-.2,-1.2){$4$}\put(1.5,-1.2){$3$}\put(3.2,-1.2){$2$}\put(4.9,-1.2){$1$}
\put(9.7,-1.2){$4$}\put(11.4,-1.2){$3$}\put(13.1,-1.2){$2$}\put(14.8,-1.2){$1$}
\put(19.7,-1.2){$4$}\put(21.4,-1.2){$3$}\put(23.1,-1.2){$2$}\put(24.8,-1.2){$1$}
\end{pspicture}
\end{center}
\caption{The linear representation of the contractions of the word
$bbb^{\dagger}b^{\dagger}$}%
\label{falt}%
\end{figure}
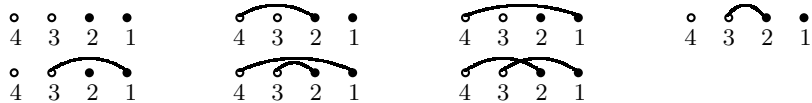

Contractions can be depicted with diagrams called linear representations. Let
us consider a word $\pi$ on the alphabet $\{b,b^{\dag}\}$ of length $n$, i.e.,
$\pi=\pi_{n}\cdots\pi_{1}$ with $\pi_{i}\in\{b,b^{\dag}\}$. We draw $n$
vertices, say $1,2,\ldots,n$, on a horizontal line, such that the point $i$
corresponds to the letter $\pi_{i}$; we represent each $b$ by a white vertex
and each letter $b^{\dagger}$ by a black vertex; a black vertex $i$ can be
connected by an undirected edge $(i,j)$ to a white vertex $j$ (but there may
also be black vertices having no edge). Importantly, the edges are drawn in
the plane above the points. This is the \emph{linear representation} of a
contraction. An example is given in Figure~\ref{falt} for the word
$bbb^{\dagger}b^{\dagger}$.

Now let us generalize the definition of normal ordering as given above. For
this we consider the alphabet consisting of two letters $a$ and $a^{\dag}$ but
where now these two letters are not assumed to satisfy any commutation
relation (like, e.g., (\ref{comm})); to avoid confusion with the bosonic
operators satisfying (\ref{comm}) we have chosen different letters here. Let
$F(a,a^{\dagger})$ be a possibly infinite word on the alphabet $\{a,a^{\dagger
}\}$. We define $\mathcal{C}(F(a,a^{\dagger}))$ to be the multiset of all
words obtained by substituting $a=e$ and $a^{\dagger}=e^{\dagger}$ whenever
$a$ precedes $a^{\dagger}$; moreover, we replace any two adjacent letters $e$
and $e^{\dagger}$ with $p$.
For example, $\mathcal{C}(aaa^{\dagger}a^{\dagger})=\{aaa^{\dagger}a^{\dagger
},eae^{\dagger}a^{\dagger},eaa^{\dagger}e^{\dagger},apa^{\dagger}%
,aea^{\dagger}e^{\dagger},epe^{\dagger},eee^{\dagger}e^{\dagger}\}$, as it is
illustrated in Figure~\ref{falt}. For each word $\pi$ in $\mathcal{C}
(F(a,a^{\dagger}))$ the double dot operation of $\pi$ is defined by deleting
all letters $e$ and $e^{\dag}$ and arranging it such that all letters
$a^{\dag}$ precede the letter $a$; clearly, $:\pi: =(a^{\dagger})^{v}%
a^{u}p^{w}$ for some $u,v,w\geq0$. We now define
\begin{equation}
\label{pnormal}\mathcal{N}_{p}[F(a,a^{\dagger})]:=\sum_{\pi\in\mathcal{C}%
(F(a,a^{\dagger}))}:\pi:.
\end{equation}
For example,
\begin{align}
\mathcal{N}_{p}((a^{\dagger}a)^{3})  &  =\,:a^{\dagger}aa^{\dagger}%
aa^{\dagger}a:+:a^{\dagger}aa^{\dagger}pa:+:a^{\dagger}ea^{\dagger}%
ae^{\dagger}a:+:a^{\dagger}paa^{\dagger}a:+:a^{\dagger}ppa:\nonumber\\
&  =(a^{\dagger})^{3}a^{3}+(2p+1)(a^{\dagger})^{2}a^{2}+p^{2}a^{\dagger}a.
\label{example}%
\end{align}
Clearly $\mathcal{N}_{p}(F(a,a^{\dagger}))$ is a generalization of the
normally ordered form $\mathcal{N}(F(a,a^{\dagger}))$, namely $\mathcal{N}%
(F(a,a^{\dagger}))=\mathcal{N}_{1}(F(a,a^{\dagger}))$.

Notice that, for all $v,u\geq0$,
\[
\mathcal{N}_{p}(a^{v}(a^{\dagger})^{u})=\sum_{i=0}^{v}\left[  \binom{v}%
{i}(u-i)+(v-1+p)\binom{v-1}{i-1}\right]  (u-1)^{\underline{i-1}}(a^{\dagger
})^{u-i}a^{v-i}
\]
where $x^{\underline{i}}=x(x-1)\cdots(x-i+1)$ and $x^{\underline{-i}%
}=(x+1)(x+2)\cdots(x+i)$ with $x^{\underline{0}}=1$. In fact,
\[
\mathcal{N}_{p}(a^{v}(a^{\dagger})^{u})=a^{\dagger}\mathcal{N}_{1}%
(a^{v}(a^{\dagger})^{u-1})+(v-1+p)\cdot\mathcal{N}_{1}(a^{v-1}(a^{\dagger
})^{u-1}).
\]

\section{Some physical considerations concerning $\mathcal{N}_{p}$}

Before deriving an explicit expression for $\mathcal{N}_{p}((a^{\dagger}%
a)^{n})$ in the next section, we want to discuss some more physical aspects of
the generalized normal ordering. Recall that in the usual case the commutation
relations of the bosonic operators are used to derive the normal ordered form
of a function of annihilation and creation operators in the form of
\textit{Wick's theorem}, cf. p.159 in \cite{BS} (there also exists a version
for fermionic operators \cite{BS}). In our case this means the following:
given the commutation relation (\ref{comm}), one may \textit{derive} the
prescription (\ref{normal}) for normal ordering. Now assume that we did not
know that the operators satisfy (\ref{comm}) - but we did know the
prescription (\ref{normal}) for arbitrary words in $b$ and $b^{\dag}$. Could
we conclude that (\ref{comm}) is satisfied? As a first step, we would find
that $\mathcal{N}([b,b^{\dag}]-I)=0$. However, to be able to conclude that
$[b,b^{\dag}]=I$, we would have to make some further assumptions.

Let us now turn to the generalized normal ordering $\mathcal{N}_{p}$. Above we
have only spoken about the letters $a$ and $a^{\dag}$ (to which $\mathcal{N}%
_{p}$ is applied) and have not interpreted these as bosonic operators. The
reason for this is simple: the prescription (\ref{pnormal}) is not consistent
with (\ref{comm})! Since this is an important point we will discuss it
explicitly. We consider the simplest case where - apart from the null
contraction - only one contraction is involved, namely the word $aa^{\dag}$.
From the definition (\ref{pnormal}) it follows that
\begin{equation}
\mathcal{N}_{p}(aa^{\dag})=:aa^{\dag}:+:pI:=a^{\dag}a+pI \label{res}%
\end{equation}
(for ease of comparison we have written the identity explicitly). On the other
hand, assuming that (\ref{comm}) holds we write $aa^{\dag}=a^{\dag}a+I$ and -
using linearity of $\mathcal{N}_{p}$ - find that
\[
\mathcal{N}_{p}(aa^{\dag})=\mathcal{N}_{p}(a^{\dag}a+I)=\mathcal{N}%
_{p}(a^{\dag}a)+\mathcal{N}_{p}(I)=a^{\dag}a+I,
\]
which clearly contradicts (for the case $p\neq1$ we are interested in) the
result (\ref{res}) obtained directly from the definition. This shows that
\textit{the letters $a$ and $a^{\dag}$ - which are normal ordered according to
$\mathcal{N}_{p}$ - cannot satisfy (\ref{comm}), \emph{i.e.}, they cannot be
interpreted straightforwardly as bosonic annihilation and creation operators!}
Note that (\ref{res}) can also be written as
\[
\mathcal{N}_{p}([a,a^{\dag}]-pI)=0.
\]
This calculation suggests that the effect of introducing the generalized
normal ordering $\mathcal{N}_{p}$ on the otherwise arbitrary letters $a$ and
$a^{\dag}$ might be equivalent to the conventional normal ordering
$\mathcal{N}$ where the letters satisfy the \textquotedblleft generalized
Heisenberg algebra\textquotedblright\ $[a,a^{\dag}]=pI$. However, this cannot
be the case since a rescaling of the letters by a factor $\sqrt{p}^{-1}$ would
result in the conventional Heisenberg algebra (\ref{comm}).

As another example consider the $q$-deformed boson operators $a_{q}$ and
$a_{q}^{\dag}$ satisfying $a_{q}a_{q}^{\dag}-qa_{q}^{\dag}a_{q}=I$, or
\begin{equation}
a_{q}a_{q}^{\dag}=qa_{q}^{\dag}a_{q}+I \label{qcomm}%
\end{equation}
(for combinatorial aspects of normal ordering $q$-deformed bosons see
\cite{Sch} and the references therein). Due to the appearing factor of $q$ it
is clear (same argument as above) that the generalized normal ordering
$\mathcal{N}_{p}$ cannot be equivalent to the standard normal ordering
$\mathcal{N}$, where the letters satisfy the $q$-deformed Heisenberg algebra
(\ref{qcomm}). Note that choosing $q=-1$ in (\ref{qcomm}) yields the canonical
anticommutation relations of fermionic annihilation and creation operators;
thus $\mathcal{N}_{p}$ is neither interpretable in terms of conventional
fermionic annihilation and creation operators $f$ and $f^{\dag}$ satisfying
$ff^{\dag}+f^{\dag}f=I$.

Let us discuss another aspect of the letters $a$ and $a^{\dag}$. Suppose we
want to represent them as operators in a Fock space. For this we first
introduce the vacuum $|0\rangle$ and define the states $|n\rangle$ through the
action of $a^{\dag}$ in the usual fashion, \emph{i.e.}, $|n\rangle
:=\frac{(a^{\dag})^{n}}{\sqrt{n!}}|0\rangle$; note that this implies $a^{\dag
}|n\rangle=\sqrt{n+1}|n+1\rangle$. The Fock space is then the linear hull of
the states $|n\rangle$. We may also define \textquotedblleft coherent
states\textquotedblright\ associated to $\gamma\in\mathbf{C}$ in the usual
manner, \emph{i.e.},
\begin{equation}
|\gamma\rangle=e^{-|\gamma|^{2}/2}\sum_{n\geq0}\frac{\gamma^{n}}{\sqrt{n!}%
}|n\rangle. \label{coher}%
\end{equation}
But now comes the really difficult part: We have to define an action of $a$ on
the states $|n\rangle$ (or $|\gamma\rangle$) which is compatible with the
generalized normal ordering $\mathcal{N}_{p}$. Assuming that $a$ destroys one
\textquotedblleft quantum\textquotedblright\ we may write $a|n\rangle
=a_{n}|n-1\rangle$ and have to determine the coefficients $a_{n}$ (we clearly
have also $a|0\rangle=0$). Note that choosing $a_{n}=\sqrt{n}$ yields the
usual Fock representation, where we can derive $a^{\dag}a=N$ (where we define
the number operator $N$ by its action $N|n\rangle=n|n\rangle$ on the states)
as well as $aa^{\dag}=N+1$ and, therefore, $[a,a^{\dag}]=I$. As we have
discussed above, this choice is not consistent with $\mathcal{N}_{p}$.
Equivalently, if one introduces first $N$ and demands $N=a^{\dag}a$ then one
also finds $a_{n}=\sqrt{n}$ and runs into the same problem. Thus, the
definition of the action $b$ on the states $|n\rangle$ seems to be rather
subtle. Note that if we want to follow the standard path we should also
introduce the dual states $\langle m|$ to be able to calculate expectation
values and also have that $a$ and $a^{\dag}$ are adjoint operators! Let us
move to the action of $a$ on the coherent states (\ref{coher}). Recall that in
the standard bosonic case one of the defining properties of the coherent
states is that they are eigenstates for the annihilation operator, i.e.,
$b|\gamma\rangle_{b}=\gamma|\gamma\rangle_{b}$ (here we have added a subscript
\textquotedblleft$b$\textquotedblright\ to indicate that the coherent states
are defined with the help of $b^{\dag}$). If we demand the analogous property
for $a$, i.e., $a|\gamma\rangle=\gamma|\gamma\rangle$ - even without knowing
the action of $a$ on the $|n\rangle$ - we are led to the conclusion
$a|n\rangle=\sqrt{n}|n-1\rangle$. However, we have discussed above that this
is not compatible with $\mathcal{N}_{p}$. Thus, it seems to be nontrivial to
generalize the usual Fock representation (and the coherent states) of the
bosonic operators $b$ and $b^{\dag}$ satisfying (\ref{comm}) and $\mathcal{N}$
to the letters $a$ and $a^{\dag}$ satisfying $\mathcal{N}_{p}$.

To conclude, it does not seem to be straightforward to give the letters $a$
and $a^{\dag}$ an interpretation as annihilation and creation operators of
some sort of particles (which are definitely not conventional bosons since
they do not satisfy (\ref{comm})). This is the reason why we speak in the
following about the \textit{letters} (and not operators) $a$ and $a^{\dag}$.
However, it would clearly be interesting to find out which algebraic relations
the letters $a$ and $a^{\dag}$ can satisfy - as generalization of (\ref{comm})
- without contradicting (\ref{pnormal}); it would be even more interesting to
find a relation from which (\ref{pnormal}) could be \textit{derived} as
consequence - in analogy to the conventional case where the usual normal
ordering can be derived from the canonical commutation relation (cf. the
discussion at the beginning of this section).

\section{An explicit formula for $\mathcal{N}_{p}((a^{\dagger}a)^{n})$}


The goal of this section to find a general formula for $\mathcal{N}%
_{p}((a^{\dagger}a)^{n})$. First of all we need the following definition. Let
$\mathcal{F}_{n}$ to be the set of all vectors $(\pi(1),\pi(2),\ldots,\pi(n))$
of length $n$ such that $\pi(i)\in\{e,i,i+1,\ldots,n\}$ and if $\pi
(i),\pi(j)\neq e$ then $\pi(i)\neq\pi(j)$. For example, the set $\mathcal{F}%
_{2}$ consists of five vectors $(e,e)$, $(e,2)$, $(1,e)$, $(2,e)$ and $(1,2)$.
We refine the set $\mathcal{F}_{n}$ denoting by $\mathcal{F}_{n,k}$ the set of
all vectors in $\mathcal{F}_{n}$ having exactly $k$ coordinates $e$.

A \emph{partition} $\pi$ of $[n]=\{1,2,\ldots,n\}$ is a collection
$P_{1},P_{2},\ldots,P_{k}$ of nonempty disjoint subsets of $[n]$, called
\emph{blocks}, such that $P_{1}\cup P_{2}\cup\cdots\cup P_{k}=[n]$. We may
assume that $P_{1},P_{2},\ldots,P_{k}$ are listed in the increasing order of
the blocks' cardinalities and we write $P_{1}<P_{2}<\cdots<P_{k}$. The set of
all partitions of $[n]$ with $k$ blocks is denoted by $P_{n,k}$. The
cardinality of $P_{n,k}$ is the well-known Stirling number $S(n,k)$ of the
second kind \cite{S1}.

The next observation show that the elements of $\mathcal{F}_{n-1,k-1}$ are
enumerated by the Stirling number $S(n,k)$.

\begin{lemma}
\label{lem1} There exists a bijection between the set $\mathcal{F}_{n-1,k-1}$
and the set $P_{n,k}$. In particular, the cardinality of the set
$\mathcal{F}_{n-1,k-1}$ is given by $S(n,k)$, the Stirling number of the
second kind.
\end{lemma}

\begin{proof}
Let $(\pi(1),\ldots,\pi(n-1))$ be any vector in $\mathcal{F}_{n-1,k-1}$ and
let $\pi$ be the vector $(\pi(1),\ldots,\pi(n-1),e)$. We define the first
block of the partition of $[n]$ by
\[
P_{1}=\{\beta_{11}=1,\beta_{12}=1+\pi(\beta_{11}),\ldots,\beta_{1\ell}%
=1+\pi(\beta_{1(\ell-1)})\},
\]
where $\pi(\beta_{1\ell})=e$, and we define the $i$-th, $i=1,2,\ldots,k$,
block of the partition of $[n]$ by
\[
P_{i}=\{\beta_{i1}=a_{i},\beta_{i2}=1+\pi(\beta_{i1}),\ldots,\beta_{i\ell
}=1+\pi(\beta_{i(\ell-1)})\},
\]
where $\pi(\beta_{i\ell})=e$ and $a_{i}$ is the minimal coordinate of the
vector $\pi$ such that $a_{i}\notin P_{j}$ for all $j=1,2,\ldots,i-1$. For
instance, if $\pi=(1,e,3,e,e)$ then $P_{1}=\{1,2\}$, $P_{2}=\{3,4\}$, and
$P_{5}=\{5\}$. From the above construction, we see that $(\pi(1),\ldots
,\pi(n-1))$ is a vector in $\mathcal{F}_{n-1,k-1}$ if and only if $P_{1}%
,P_{2},\ldots,P_{k}$ is a partition of $[n]$, as required.
\end{proof}

\bigskip

For example, Lemma~\ref{lem1} for $n=2$ gives
\[
\rho(ee)=\{1\},\{2\},\{3\};\,\rho(e2)=\{1\},\{23\};\,\rho
(1e)=\{12\},\{3\};\,\rho(12)=\{123\};\,\rho(2e)=\{13\},\{2\}.
\]

Now we prove a bijection between the set of contractions of $(a^{\dagger
}a)^{n}$ and the set $\mathcal{F}_{n}$. To do that we need the following
definition. We say that a contraction $\pi\in\mathcal{C}((a^{\dagger}a)^{n})$
\emph{has degree }$k$ if $:\pi:=(a^{\dagger})^{v}a^{w}p^{k}$ for $v,w\geq0$.
The set of the contractions in $\mathcal{C}((a^{\dagger}a)^{n})$ having degree
$k$ is denoted by $C_{n,k}$.

\begin{lemma}
\label{lem2} There exists a bijection between the set of contractions
$C_{n,k-1}$ and the set $\mathcal{F}_{n-1,k-1}$.
\end{lemma}

\begin{proof}
Let $w=w_{2n}w_{2n-1}\ldots w_{1}$ be any contraction of $(a^{\dagger}a)^{n}$.
For each $j=1,2,\ldots,n-1$, define $\pi(j)=e$ if $w_{2j}=a^{\dagger}$, and
$\pi(j)=i$ if $w_{2j}=e^{\dagger}$ and $w_{2i+1}=e$, where $i$ is minimal and
greater than $j$. The definition of $\pi$ implies that $w$ is a contraction of
$(a^{\dagger}a)^{n}$ if and only if the vector $\pi=(\pi(1),\pi(2),\ldots
,\pi(n-1))\in\mathcal{F}_{n-1}$. Moreover, if $w$ is a contraction then the
vector $\pi$ has $k-1$ coordinates $e$ if and only if $k-1=\#\{j|w_{2j}%
=a^{\dagger}\}$, or, in other words, $\pi\in\mathcal{F}_{n-1,k-1}$ if and only
if $w\in C_{n,k-1}$.
\end{proof}

\bigskip

Let $Q$ be any subset of $[n]=\{1,2\ldots,n\}$. We say that $Q$ has a
\emph{rise} at $i$ if $i,i+1\in Q$. The number of rises of $Q$ is denoted by
$rise(Q)$. Let $\pi=P_{1},P_{2},\ldots,P_{k}$ be any partition of $[n]$, we
define $rise(\pi)=\sum_{j=1}^{k}rise(P_{j})$ and we say that $\pi$ has exactly
$rise(\pi)$ rises. Lemma~\ref{lem1} together with Lemma~\ref{lem2} give then
the following result.

\begin{proposition}
\label{pro3} There exists a bijection between the set of contractions
$C_{n,k}$ and the set of partitions of $n$ with $k$ rises.
\end{proposition}

We are now ready to give a formula for $\mathcal{N}_{p}((a^{\dagger}a)^{n})$.

\begin{theorem}
\label{mth1} For all $n\geq1$,
\begin{equation}
\label{stirlp}\mathcal{N}_{p}((a^{\dagger}a)^{n})=\sum_{k=0}^{n}%
S_{p}(n,k)(a^{\dagger})^{k}a^{k},
\end{equation}
where $S_{p}(n,k)$ satisfies the following recurrence relation
\[
S_{p}(n,k)=(k-1+p)S_{p}(n-1,k)+S_{p}(n-1,k-1),
\]
with the initial conditions $S_{p}(n,1)=p^{n-1}$ and $S_{p}(n,k)=0$ for all
$k>n$.
\end{theorem}

\begin{proof}
Denote by $S(n,k;m)$ the number of contractions $\pi$ of $(a^{\dagger}a)^{n}$
such that $:\pi:=(a^{\dagger})^{k}a^{k}p^{m}$. From Proposition~\ref{pro3}, we
get that $S(n,k;m)$ is the number of partitions of $[n]$ into $k$ blocks
$P_{1},\ldots,P_{k}$ with $m$ rises. To find a recurrence relation for the
sequence $S(n,k;m)$, we consider the position of $n$ in the blocks
$P_{1},\ldots,P_{k}$ with $P_{1}<P_{2}<\cdots<P_{k}$. If $P_{k}=\{n\}$ then
there are $S(n-1,k-1;m)$ such partitions$;$ if $n\in P_{i}$ and $n-1\not \in
P_{i}$ then there are $S(n-1,k;m)$ partitions; if $n,n-1,\in P_{i}$ then the
number of partitions is $S(n-1,k;m-1)$. Therefore,
\[
S(n,k;m)=(k-1)S(n-1,k;m)+S(n-1,k;m-1)+S(n-1,k-1;m),
\]
which is equivalent to
\[
S_{p}(n,k)=(k-1+p)S_{p}(n-1,k)+S_{p}(n-1,k-1),
\]
where $S_{p}(n,k)=\sum_{m=0}^{n-k}S(n,k;m)p^{m}$. Hence,
\[
\mathcal{N}_{p}((a^{\dagger}a)^{n})=\sum_{k=0}^{n}\sum_{m=0}^{n-k}%
S(n,k;m)p^{m}(a^{\dagger})^{k}a^{k}=\sum_{k=0}^{n}S_{p}(n,k)(a^{\dagger}%
)^{k}a^{k}.
\]
The initial conditions can be checked directly from the definitions.
\end{proof}

\bigskip

As we mentioned above the polynomials $S_{p}(n,k)$ satisfy the recurrence
relation
\[
S_{p}(n,k)=(k-1+p)S_{p}(n-1,k)+S_{p}(n-1,k-1),
\]
with the initial conditions $S_{p}(n,1)=p^{n-1}$ and $S_{p}(n,k)=0$ for all
$k>n$. Considering $p=1$ yields the recurrence relation (\ref{Stirl}) of the
conventional Stirling numbers $S_{1}(n,k)\equiv S(n,k)$. If we define
\[
S_{p}(x;k)=\sum_{n\geq k}S_{p}(n,k)\frac{x^{n}}{n!},
\]
it is not hard to see that
\[
S_{p}(x;k)=\frac{(p-1)!}{(p+k-1)!}+\frac{e^{px}}{(k-1)!}\sum_{j=0}^{k-1}%
\frac{(-1)^{k-1-j}\binom{k-1}{j}}{p+j}e^{jx}=\int_{0}^{x}\frac{e^{pt}}%
{(k-1)!}(e^{t}-1)^{k-1}dt.
\]
Thus the generating function
\[
S_{p}(x,y)=\sum_{n\geq1}\sum_{k=1}^{n}S_{p}(n,k)y^{k}\frac{x^{n}}{n!}%
\]
is given by
\[
S_{p}(x,y)=\int_{0}^{x}ye^{pt}e^{y(e^{t}-1)}dt.
\]
Notice that
\[
S_{1}(x,y)=e^{y(e^{x}-1)}
\]
is the generating function for the Stirling numbers of the second kind. One of
the first explicit expressions for the numbers $S_{p}(n,k)$ was given by
d'Ocagne~\cite{O}:
\[
S_{p}(n,k)=\frac{(-1)^{k-1}}{(k-1)!}\sum_{j=0}^{k-1}(-1)^{j}\binom{k-1}%
{j}(p+j)^{n-1},
\]
for all $n\geq k\geq1$. With this expression, Theorem~\ref{mth1} gives an
explicit formula for $\mathcal{N}_{p}((a^{\dagger}a)^{n})$.

\begin{theorem}
\label{mth2} For all $n\geq1$,
\[
\mathcal{N}_{p}((a^{\dagger}a)^{n})=\sum_{k=0}^{n}\left(  \frac{(-1)^{k-1}%
}{(k-1)!}\sum_{j=0}^{k-1}(-1)^{j}\binom{k-1}{j}(p+j)^{n-1}\right)
(a^{\dagger})^{k}a^{k}.
\]

\end{theorem}

Theorem \ref{mth2} gives equation \eqref{eqkat} for $p=1$. For $p=0$,
Theorem~\ref{mth1} gives
\[
\mathcal{N}_{0}((a^{\dagger}a)^{n})=\sum_{j=0}^{n}S_{0}(n,j)(a^{\dagger}%
)^{j}a^{j},
\]
where $S_{0}(n,k)$ satisfies the following recurrence relation $S_{0}%
(n,k)=(k-1)S_{0}(n-1,k)+S_{0}(n-1,k-1)$ with initial condition $S_{0}%
(n,1)=\delta_{n,1}$ and $S_{0}(n,k)=0$, for $k>n$. Thus, $\sum_{k=1}^{n}%
S_{0}(n,k)=B_{n-1}$, the $(n-1)$-th Bell number.

\section{A general class of normal orderings}

In this section we want to outline a rather general approach to
generalizations of the normal ordering using contractions and the double dot
operation. As in the definition of $\mathcal{N}_{p}$ in Section II, we
consider words $F(a,a^{\dag})$ in the letters $a$ and $a^{\dag}$ (which
satisfy no relation). The set of \textit{resulting contractions} - in the
following abbreviated by \textit{r-contractions} - of $F(a,a^{\dag})$ will be
denoted by $\mathcal{RC}(F(a,a^{\dag}))$ and it is the multiset of all words
which result by pairing the letters $a^{\dag}$ with the letters $a$ and
omitting them, see Figure~\ref{falt}. For example, if $F(a,a^{\dag})=aa^{\dag
}aa^{\dag}$ then $\mathcal{RC}(aa^{\dag}aa^{\dag})=\{aa^{\dag}aa^{\dag
},aa^{\dag},aa^{\dag},aa^{\dag},1,1\}$, where the first r-contraction comes
from the null contraction, the next r-contraction results by pairing the first
$a$ with the first $a^{\dag}$, the next r-contraction results by pairing the
first $a$ with the second $a^{\dag}$, the next r-contraction results by
pairing the second $a$ with the second $a^{\dag}$, the next r-contraction
results by pairing the first $a$ with the first $a^{\dag}$ \textit{and} the
second $a$ with the second $a^{\dag}$, and the final r-contraction results by
pairing the first $a$ with the second $a^{\dag}$ \textit{and} the second $a$
with the first $a^{\dag}$.

Since each r-contraction $\pi$ contains only the letters $a$ and $a^{\dag}$ we
can define $:\pi:$ as in the conventional case by arranging the letters so
that all $a^{\dag}$ precede all $a$ (together with $:1:=1$). Now, assume that
we have a prescription associating to each $\pi\in\mathcal{RC}(F(a,a^{\dag}))$
a \textit{weight} $\mathcal{W}(\pi)$ (concrete examples will follow soon).
Then we can define the $\mathcal{W}$-\textit{generalized normal ordering} by
\begin{equation}
\mathcal{N}_{\mathcal{W}}(F(a,a^{\dag})):=\sum_{\pi\in\mathcal{RC}%
(F(a,a^{\dag}))}\mathcal{W}(\pi):\pi:. \label{generalnorm}%
\end{equation}
This generalizes the normal ordering (\ref{normal}) and also (\ref{pnormal}).
Note that there exists a bijection between contractions and r-contractions;
however, in our above definition of contractions in Section II we have
implicitly built in the weights. To separate these aspects we have introduced
here in the general case the r-contractions. Clearly, if the weights of all
r-contractions are equal to one then (\ref{generalnorm}) reduces to
(\ref{normal}) and one may identify contractions with r-contractions.

Let us define the \textit{r-degree} of a r-contraction as the number of pairs
of letters $a$ and $a^{\dag}$ which have been contracted in the original word.
We furthermore define a \textit{crossing} of a contraction as in \cite{MS},
i.e., if $e=(i,j)$ and $f=(k,l)$ are two edges in the word $F(a,a^{\dag})$
then we say that $e$ \textit{crosses} $f$ if they intersect with each other,
i.e., if $i<k<j<l$ or $k<i<l<j$(e.g., only the last contraction of
Figure~\ref{falt} is crossing). A contraction having no crossings is called
\textit{noncrossing}. Clearly, all contractions of degree less than or equal
to one are noncrossing. The \textit{crossing number} $\mathcal{X}(\pi)$ of a
contraction $\pi$ is defined to be the number of crossings and the crossing
number of a r-contraction is defined to be the crossing number of the
associated contraction.

For the general class of normal orderings we are interested in we require the
weights of the r-contractions to satisfy the following properties:

\begin{enumerate}
\item The weight is the product of a \textit{crossing weight} and a
\textit{contraction weight}, where the crossing weight only depends on the
number of crossings and the contraction weight does not depend on the number
of crossings.

\item The contraction weight $\mathcal{CW}$ of a r-contraction of r-degree one
depends only on the distance of the letters which are contracted. The
contraction weight of a r-contraction of r-degree $k$ is the product of the
weights of the $k$ ``subcontractions'' it consists of.

\item The crossing weight of a r-contraction $\pi$ is given by $q^{\mathcal{X}%
(\pi)}$, where $q$ is the crossing weight of a r-contraction with one crossing.
\end{enumerate}

To formalize these rules we introduce an infinite sequence $\omega$ of
weights, that is $\omega=(\omega_{-1},\omega_{0}=1,\omega_{1},\omega
_{2},\ldots)$. The elements have the following interpretation: $\omega_{-1}$
is the crossing weight for one crossing denoted by $q$ above, $\omega_{0}=1$
is the weight of the null contraction and $\omega_{n}$ with $n\geq1$ are the
contraction weights for a r-contraction of degree one where the pair of
contracted letters has distance $n$. We are now ready to define the $\omega
$-\textit{generalized normal ordering} by letting
\begin{equation}
\mathcal{N}_{\omega}(F(a,a^{\dag})):=\sum_{\pi\in\mathcal{RC}(F(a,a^{\dag}%
))}\omega_{-1}^{\mathcal{X}(\pi)}\mathcal{CW}_{\omega}(\pi):\pi:.
\label{omeganorm}%
\end{equation}
Comparing (\ref{omeganorm}) with (\ref{generalnorm}) shows that the weight of
a r-contraction $\pi$ is given in terms of $\omega$ by $\omega_{-1}%
^{\mathcal{X}(\pi)}\mathcal{CW}_{\omega}(\pi)$. Before giving explicit
examples of the sequence $\omega$ we consider some concrete examples of the
$\omega$-generalized normal ordering. As a simple example one has
$\mathcal{N}_{\omega}((a^{\dag}a)^{2})=\omega_{0}(a^{\dag})^{2}a^{2}%
+\omega_{1}a^{\dag}a$ (where we have explicitly denoted the weight of the null
contraction with $\omega_{0}=1$). A slightly more involved example is given
by
\begin{equation}
\mathcal{N}_{\omega}((a^{\dag}a)^{3})=\omega_{0}(a^{\dag})^{3}a^{3}%
+\{2\omega_{1}+\omega_{3}\}(a^{\dag})^{2}a^{2}+\omega_{1}^{2}a^{\dag}a.
\label{ex1}%
\end{equation}
Clearly, if the sequence $\omega$ is given by $\omega_{1}=p$ and $\omega
_{i}=1$ for $i\neq1$ then (\ref{ex1}) reduces to (\ref{example}). Note that up
to now no r-contractions with crossings have appeared. In the next example
there appears one crossing (when the first $a$ is contracted with the third
$a^{\dag}$ and the second $a$ with the fourth $a^{\dag}$):
\begin{equation}
\mathcal{N}_{\omega}((a^{\dag}a)^{4})=\omega_{0}(a^{\dag})^{4}a^{4}%
+\{3\omega_{1}+2\omega_{3}+\omega_{5}\}(a^{\dag})^{3}a^{3}+\{2\omega_{1}%
^{2}+2\omega_{1}\omega_{3}+\omega_{-1}\omega_{3}^{2}+\omega_{1}\omega
_{5}\}(a^{\dag})^{2}a^{2}+\omega_{1}^{3}a^{\dag}a. \label{ex2}%
\end{equation}
Let us now consider some special sequences $\omega$:

\begin{enumerate}
\item $\omega^{(1)}=(1,1,1,1,\ldots)$: The weight of all r-contractions is one
(independent of crossings). This corresponds to the conventional normal
ordering, i.e., $\mathcal{N}_{\omega^{(1)}}=\mathcal{N}_{1}=\mathcal{N}$.

\item $\omega^{(p)}=(1,1,p,1,\ldots)$: Since $\omega_{-1}=1$ crossings play no
role. Only $\omega_{1}=p$ is not equal to one, thus all r-contractions of
degree one have weight one except those where the contracted operators are
adjacent - then the r-contraction has weight $p$. This reproduces exactly the
generalized normal ordering $\mathcal{N}_{p}$ from above, i.e., $\mathcal{N}%
_{\omega^{(p)}}=\mathcal{N}_{p}$.

\item $\omega^{nc}=(0,1,1,1,\ldots)$: All r-contractions have weight one,
except those, where at least one crossing occurs - these r-contractions have
weight zero. Thus, this example yields the \textit{noncrossing normal
ordering} considered in \cite{MS}.
\end{enumerate}

Returning to the general case, one may consider again the particular word
$F(a,a^{\dag})=(a^{\dag}a)^{n}$. Denoting the set of r-contractions of
r-degree $l$ by $\mathcal{RC}^{(l)}((a^{\dag}a)^{n})$, we have the disjoint
union $\mathcal{RC}((a^{\dag}a)^{n})=\bigcup_{l=0}^{n}\mathcal{RC}%
^{(l)}((a^{\dag}a)^{n})$. Note that for any $\pi\in\mathcal{RC}^{(l)}%
((a^{\dag}a)^{n})$ there have been $l$ pairs of letters $a$ and $a^{\dag}$
contracted, leaving $n-l$ letters $a$ and $a^{\dag}$. Thus, $\pi
\in\mathcal{RC}^{(l)}((a^{\dag}a)^{n})\Rightarrow:\pi:=(a^{\dag})^{n-l}%
a^{n-l}$. Combining these results yields
\begin{equation}
\mathcal{N}_{\omega}((a^{\dag}a)^{n})=\sum_{l=0}^{n}\left\{  \sum_{\pi
\in\mathcal{RC}^{(l)}((a^{\dag}a)^{n})}\omega_{-1}^{\mathcal{X}(\pi
)}\mathcal{CW}_{\omega}(\pi)\right\}  (a^{\dag})^{n-l}a^{n-l}. \label{stir}%
\end{equation}
Thus, introducing the $\omega$-\textit{generalized Stirling numbers of second
kind} by
\begin{equation}
S_{\omega}(n,k):=\sum_{\pi\in\mathcal{RC}^{(n-k)}((a^{\dag}a)^{n})}\omega
_{-1}^{\mathcal{X}(\pi)}\mathcal{CW}_{\omega}(\pi), \label{stirlo}%
\end{equation}
we can write (\ref{stir}) in analogy to (\ref{eqkat}) and (\ref{stirlp}) as
\begin{equation}
\mathcal{N}_{\omega}((a^{\dag}a)^{n})=\sum_{k=1}^{n}S_{\omega}(n,k)(a^{\dag
})^{k}a^{k}.
\end{equation}
The usual Stirling numbers are obtained if all weights are equal to one,
\emph{i.e.}, $\omega=\omega^{(1)}$ from above:
\[
S(n,k)=S_{\omega^{(1)}}(n,k)=|\mathcal{RC}^{(n-k)}((a^{\dag}a)^{n})|.
\]
If $\omega=\omega^{(p)}$ from above then $S_{\omega^{(p)}}(n,k)=S_{p}(n,k)$.
For the general Stirling numbers (\ref{stirlo}) it might be interesting to
obtain some general results (\emph{e.g.}, explicit values, recurrence
relation). Let us mention some simple observations. Since there exists exactly
one null contraction, $S_{\omega}(n,n)=\omega_{0}=1$. On the other hand, there
exists exactly one r-contraction of r-degree $n-1$, where all letters except
the \textquotedblleft boundaries\textquotedblright\ are contracted (thus, all
adjacent pairs $aa^{\dag}$ are contracted), implying $S_{\omega}%
(n,1)=\omega_{1}^{n-1}$. These two values can be seen nicely in the above
explicit examples (\ref{ex1}) and (\ref{ex2}). Of course, one may define in
analogy to the usual case also $\omega$-\textit{generalized Bell numbers} by
$B_{\omega}(n)=\sum_{k=0}^{n}S_{\omega}(n,k)$.

\section{Conclusion}

We have generalized results of \cite{Kat02} by defining a \emph{refined}
version of the normal ordering, and we have discussed the physical aspects of
the approach considered. The essence of the generalization $\mathcal{N}_{p}$
can be described as follows. For a given expression $F(a^{\dagger},a)$, we use
the set of all contractions in order to get the normally ordered form
$\mathcal{N}(F(a^{\dagger},a))$. Each contraction can be seen as having weight
one (that is all contractions are equidistributed). We associate a weight to
each contraction (corresponding to the number of $ee^{\dagger}$'s in the
expression). In this way, we can study the normally ordered form for the given
expression based on the set of contractions with a given weight.

Several authors studied the standard normally ordered form of different kind
of expressions, such as $(a^{\dagger}a)^{n}$, $(a^{r}+a^{\dagger})^{n}$ and
$(a+(a^{\dagger})^{r})^{n}$ \cite{thesis}. Putting on the side a potential
operational value, the form $\mathcal{N}_{p}(F(a,a^{\dagger}))$ is interesting
because it says more than the standard normally ordered form $\mathcal{N}%
_{1}(F(a,a^{\dagger}))$. Indeed, this generalization implies extra information
on the set of the contractions of a given expression $F(a,a^{\dagger})$.
Additionally, we have introduced a further generalization, where contractions
between vertices of different distances are allowed to have arbitrary weights.


\bigskip

\textbf{Acknowledgement}. We would like to thank Jacob Katriel for his
encouragement and interest in this work, Ed Corrigan and Chris Fewster for
helpful discussion. SS is supported by EPSRC of the U.K. The comment of the
anonymous referees helped to greatly improve the paper.

\end{document}